\documentclass[letterpaper,conference]{IEEEtran}
\IEEEoverridecommandlockouts
\usepackage[utf8]{inputenc}
\usepackage[numbers]{natbib}
\usepackage{graphicx}
\usepackage[hyphens]{url}
\usepackage[colorlinks,citecolor=blue,urlcolor=blue,linkcolor=blue,bookmarks=false,hypertexnames=true]{hyperref}
\usepackage{pifont}
\usepackage{float}
\usepackage{threeparttable}
\usepackage[ruled,vlined,linesnumbered]{algorithm2e}
\usepackage{algpseudocode}
\usepackage{subcaption}
\usepackage[bottom,multiple,para]{footmisc}
\usepackage[left=0.625in, right=0.625in, top=0.75in, bottom=1in]{geometry}

\newcommand{\cmark}{\ding{51}}
\newcommand{\xmark}{\ding{55}}

\DeclareRobustCommand\optionalsec[1]{%
  \ifnum\pdfstrcmp{#1}{\thesection}=0\else#1.\fi
}
  
\newcommand{\Desc}[2]{\makebox[2em][l]{#1}#2}

\begin{document}

\title{Distributed Attestation Revocation in Self-Sovereign Identity
\thanks{This work was funded by NWO/TKI grant BLOCK.2019.004.}
}

\author{
\IEEEauthorblockN{Rowdy Chotkan}
\IEEEauthorblockA{\textit{Distributed Systems} \\
\textit{Delft University of Technology}\\
Delft, The Netherlands \\
R.M.Chotkan-1@tudelft.nl}
\and
\IEEEauthorblockN{Jérémie Decouchant}
\IEEEauthorblockA{\textit{Distributed Systems} \\
\textit{Delft University of Technology}\\
Delft, The Netherlands \\
J.Decouchant@tudelft.nl}
\and
\IEEEauthorblockN{Johan Pouwelse}
\IEEEauthorblockA{\textit{Distributed Systems} \\
\textit{Delft University of Technology}\\
Delft, The Netherlands \\
J.A.Pouwelse@tudelft.nl}
}

\maketitle
\thispagestyle{plain}
\pagestyle{plain}

\begin{abstract}
    Self-Sovereign Identity (SSI) aspires to create a standardised identity layer for the Internet by placing citizens at the centre of their data, thereby weakening the grip of big tech on current digital identities. However, as millions of both physical and digital identities are lost annually, it is also necessary for SSIs to possibly be revoked to prevent misuse. Previous attempts at designing a revocation mechanism typically violate the principles of SSI by relying on central trusted components. This lack of a distributed revocation mechanism hampers the development of SSI. In this paper, we address this limitation and present the first fully distributed SSI revocation mechanism that does not rely on specialised trusted nodes. Our novel gossip-based propagation algorithm disseminates revocations throughout the network and provides nodes with a proof of revocation that enables offline verification of revocations. We demonstrate through simulations that our protocol adequately scales to national levels.
\end{abstract}

\begin{IEEEkeywords}
Self-Sovereign Identity, revocation, offline verification
\end{IEEEkeywords}

\bstctlcite{IEEEexample:BSTcontrol}
    
\section{Introduction}
In our modern societies, citizens do not own their identities. The European Union recently announced that it would maintain a trusted and secure digital identity for each European citizen~\cite{european_commission_2021_eid}. The majority of current digital identities are also maintained by Big Tech, which results in potential privacy issues as the digital presence of citizens can be monitored~\cite{tene2012big}. Furthermore, these digital identities can be revoked at the platform owner's discretion leading to loss of access to a plethora of other dependent connected services~\cite{rogers2020deplatforming}.
The \textit{Self-Sovereign Identity} (SSI) concept overcomes these digital and societal issues by relying directly on the Internet, which currently does not embed any native method to determine who is communicating with whom~\cite{cameron2005laws}. As such, the SSI movement aims to create a standardised identity layer for the Internet, generating digital trust through verifiable identities and putting citizens at the centre of their data. 

Previous works laid out the relevant principles and architectures of SSI~\citep{Muhle2018AIdentity}. However, in particular, SSI must be able to handle compromised identities, which might appear as a consequence of theft, loss, or a data breach. For the past five years in the USA, more than a million data breaches occurred annually~\citep{johnson_2021}, resulting in the loss of billions of credentials. Furthermore, 0.8\% of UK passports~\cite{office_2018} and 340,000 identity documents in The Netherlands~\cite{proeftuin_2015} are lost annually. Revocation of these credentials is required to minimise further potential negative consequences.

Identity revocation remains a key technical challenge in Self-Sovereign Identity. As portrayed by~\autoref{tab:related_works} (further discussed in~\autoref{sec:related_work}), previous SSI distributed revocation designs paved the way but they are still incomplete. Existing SSI and digital identity solutions such as Sovrin\footnote{\url{https://sovrin.org/}}, Veramo\footnote{\url{https://veramo.io/}} (formerly known as uPort) and IRMA\footnote{\url{https://irma.app/}} violate the principles of SSI itself by addressing revocation through centralisation and trusted third  parties, whereas the cardinal requirement for SSI is an authoritarian-free ecosystem. Furthermore, recent natural disasters demonstrate that assuming the presence of always up-and-running digital infrastructure is not safe~\citep{zhang_rui_2021}. Digital identities are to be disaster-proof. Dependence on central parties for verification prevents offline usability and, moreover, introduces inherent inequalities in the network, leading to censorship or privacy issues~\cite{sovrin_2017}.

In this paper, we address the identity revocation problem and alleviate an important issue that has been hampering the mass deployment of Self-Sovereign Identities.
In a summary, we make the following contributions. We present the first revocation protocol for SSI that is fully distributed, supports offline verification of revocations and does not rely on additional trust assumptions. We evaluate this protocol using our  pioneering serverless phone-to-phone infrastructure~\citep{Halkes2011UDPWild,ZeilTech}, and a fully functioning SSI application that is backed by the Dutch government, which demonstrate the usability of distributed revocation on smartphones at a national level.

\begin{table*}[ht!]
\centering
\resizebox{\textwidth}{!}{
\begin{threeparttable}[]
\caption{Comparison of existing revocation solutions} \label{tab:related_works}

\begin{tabular}{|l|c|c|c|l|c|c|c|c|c|c|}
\hline
 &
  \textbf{\textbf{Domain}} &
  \textbf{Type} &
  \textbf{Mature\tnote{1}} &
  \multicolumn{1}{c|}{\textbf{Description}} &
  \textbf{\begin{tabular}[c]{@{}c@{}}No network \\ operators\end{tabular}} &
  \textbf{\begin{tabular}[c]{@{}c@{}}Offline \\ availability\end{tabular}} &
  \textbf{\begin{tabular}[c]{@{}c@{}}No authority \\ interactivity\end{tabular}} &
  \textbf{\begin{tabular}[c]{@{}c@{}}Offline \\ verification\end{tabular}} &
  \textbf{\begin{tabular}[c]{@{}c@{}}No \\ SPOF\end{tabular}} &
  \textbf{\begin{tabular}[c]{@{}c@{}}Full\\ accuracy\end{tabular}} \\ \hline
\textbf{This work~(\autoref{sec:revocation_design})} &
  SSI &
  Attestation &
  \Large\textcolor{green}{\cmark} &
  First fully distributed SSI revocation mechanism. &
  \Large\textcolor{green}{\cmark} &
  \Large\textcolor{green}{\cmark} &
  \Large\textcolor{green}{\cmark} &
  \Large\textcolor{green}{\cmark} &
  \Large\textcolor{green}{\cmark} &
  \Large\textcolor{green}{\cmark} \\ \hline \hline
\textbf{Abraham et al.~\citep{abraham_2020}} &
  SSI &
  Attestation &
  \textcolor{green}{\cmark} &
  Revocations stored on public permissioned blockchain. &
  \textcolor{red}{\xmark} &
  \textcolor{green}{\cmark} &
  \textcolor{green}{\cmark} &
  \textcolor{green}{\cmark} &
  \textcolor{green}{\cmark} &
  \textcolor{green}{\cmark} \\ \hline
\textbf{Baars~\citep{Baars2016TowardsTechnology}} &
  SSI &
  Credential &
  \textcolor{red}{\xmark} &
  Revocations stored on smart contracts. &
  \textcolor{red}{\xmark} &
  \textcolor{red}{\xmark} &
  \textcolor{red}{\xmark} &
  \textcolor{red}{\xmark} &
  \textcolor{red}{\xmark} &
  \textcolor{green}{\cmark} \\ \hline
\textbf{Eschenauer and Gligor~\citep{eschenauer2002key}} &
  DSN &
  Node &
  \textcolor{red}{\xmark} &
  Single authority propagates revocations. &
  \textcolor{red}{\xmark} &
  \textcolor{green}{\cmark} &
  \textcolor{red}{\xmark} &
  \textcolor{green}{\cmark} &
  \textcolor{red}{\xmark} &
  \textcolor{green}{\cmark} \\ \hline
\textbf{Haas et al.~\citep{Haas2011EfficientDistribution}} &
  VANET &
  Certificate &
  \textcolor{red}{\xmark} &
  RSUs and v2v propagation. &
  \textcolor{red}{\xmark} &
  \textcolor{green}{\cmark} &
  \textcolor{green}{\cmark} &
  \textcolor{green}{\cmark} &
  \textcolor{green}{\cmark} &
  \textcolor{red}{\xmark} \\ \hline
\textbf{IRMA~\citep{alpar2017irma}} &
  SSI &
  Attestation &
  \textcolor{green}{\cmark} &
  Uses centralised database. &
  \textcolor{red}{\xmark} &
  \textcolor{red}{\xmark} &
  \textcolor{red}{\xmark} &
  \textcolor{red}{\xmark} &
  \textcolor{red}{\xmark} &
  \textcolor{red}{\xmark} \\ \hline
\textbf{Laberteaux et al.~\citep{laberteaux2008security}} &
  VANET &
  Certificate &
  \textcolor{red}{\xmark} &
  RSUs and v2v propagation. &
  \textcolor{red}{\xmark} &
  \textcolor{green}{\cmark} &
  \textcolor{green}{\cmark} &
  \textcolor{green}{\cmark} &
  \textcolor{green}{\cmark} &
  \textcolor{green}{\cmark} \\ \hline
\textbf{Lasla et al.~\citep{Lasla2018EfficientITS}} &
  C-ITS &
  Node &
  \textcolor{red}{\xmark} &
  Revocations stored on blockchain and RSUs. &
  \textcolor{red}{\xmark} &
  \textcolor{green}{\cmark} &
  \textcolor{green}{\cmark} &
  \textcolor{red}{\xmark} &
  \textcolor{green}{\cmark} &
  \textcolor{green}{\cmark} \\ \hline
\textbf{Liau et al.~\citep{liau2005}} &
  P2P &
  Certificate &
  \textcolor{red}{\xmark} &
  Uses distribution points and P2P communication. &
  \textcolor{red}{\xmark} &
  \textcolor{green}{\cmark} &
  \textcolor{red}{\xmark} &
  \textcolor{green}{\cmark} &
  \textcolor{green}{\cmark} &
  \textcolor{green}{\cmark} \\ \hline
\textbf{Popescu et al.~\citep{propescu2003}} &
  DS &
  Certificate &
  \textcolor{red}{\xmark} &
  Revocations handled locally by authority. &
  \textcolor{green}{\cmark} &
  \textcolor{red}{\xmark} &
  \textcolor{red}{\xmark} &
  \textcolor{green}{\cmark} &
  \textcolor{red}{\xmark} &
  \textcolor{red}{\xmark} \\ \hline
\textbf{Sovrin~\citep{tobin2016inevitable}} &
  SSI &
  Attestation &
  \textcolor{green}{\cmark} &
  Uses public permissioned blockchain. &
  \textcolor{red}{\xmark} &
  \textcolor{red}{\xmark} &
  \textcolor{green}{\cmark} &
  \textcolor{red}{\xmark} &
  \textcolor{green}{\cmark} &
  \textcolor{green}{\cmark} \\ \hline
\textbf{Speelman~\citep{speelman_2020}} &
  SSI &
  Credential &
  \textcolor{red}{\xmark} &
  Uses active verification with issuer. &
  \textcolor{green}{\cmark} &
  \textcolor{red}{\xmark} &
  \textcolor{red}{\xmark} &
  \textcolor{red}{\xmark} &
  \textcolor{red}{\xmark} &
  \textcolor{green}{\cmark} \\ \hline
\textbf{Stokkink et al.~\cite{stokkink2018deployment,stokkink2020truly}} &
  SSI &
  Attestation &
  \textcolor{green}{\cmark} &
  (Central) revocation registers, DLs, validity terms. &
  \textcolor{green}{\cmark} &
  \textcolor{red}{\xmark} &
  \textcolor{red}{\xmark} &
  \textcolor{red}{\xmark} &
  \textcolor{red}{\xmark} &
  \textcolor{green}{\cmark} \\ \hline
\textbf{Veramo (uPort)~\citep{lundkvist_uport}} &
  SSI &
  Attestation &
  \textcolor{red}{\xmark} &
  Uses public permissionless blockchain. &
  \textcolor{green}{\cmark} &
  \textcolor{red}{\xmark} &
  \textcolor{green}{\cmark} &
  \textcolor{red}{\xmark} &
  \textcolor{green}{\cmark} &
  \textcolor{green}{\cmark} \\ \hline
\textbf{Xu et al.~\citep{xu2020identity}} &
  SSI &
  Node &
  \textcolor{red}{\xmark} &
  List of accepted nodes stored on blockchain. &
  \textcolor{red}{\xmark} &
  \textcolor{green}{\cmark} &
  \textcolor{green}{\cmark} &
  \textcolor{red}{\xmark} &
  \textcolor{green}{\cmark} &
  \textcolor{green}{\cmark} \\ \hline
\end{tabular}%

\begin{tablenotes}
\vspace{5pt}
\item[1] Refers to maturity of the technology.
\end{tablenotes}
\end{threeparttable}
}

\end{table*}

\section{Problem Description}
\begin{figure}[b]
    \centering
    \includegraphics[width=\linewidth]{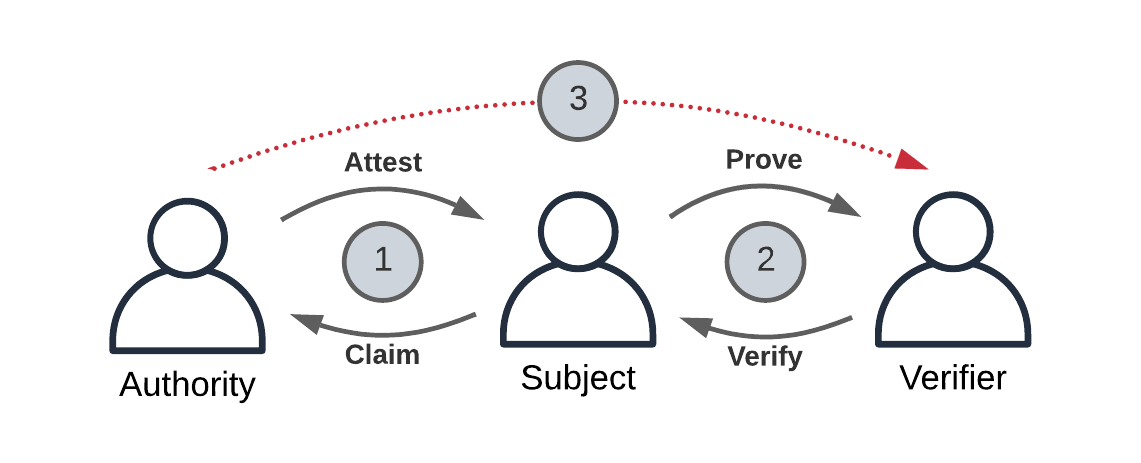}
    \caption{Interactions in a Self-Sovereign Identity system.}
    \label{fig:abstract_rev}
\end{figure}

Because of theft or loss, digital identities may become compromised. To mitigate further damage, compromised credentials must be revoked. Revocation is also required when a credential becomes (prematurely) voided, e.g., an employee who is no longer employed by a company should no longer be given access to its infrastructure.

\autoref{fig:abstract_rev} portrays the interactions between the three relevant parties of an SSI system following the definitions set out by the W3C~\cite{issuer_did}. An Issuer attests to a claim of a Subject by creating an attestation (step 1). A Verifier is then able to determine the validity of said claim by cryptographically verifying the attestation (step 2). In the instance that the attestation is to be revoked (step 3), the verification of the attestation by the Verifier must fail. The Subject can not be trusted to make the revocation available to the Verifier, as this sensibly goes against its own interest. Furthermore, revocations must be disseminated to any party that needs to verify the corresponding credential.

Bringing the Verifier in direct contact with the Issuer would go against the principles of Self-Sovereign Identity as this would defeat the purpose of attestations~\cite{Muhle2018AIdentity}. Furthermore, the cardinal requirement of SSI is that no third party is required or able to observe or otherwise interfere with the creation or verification of identity data~\citep{stokkink2020truly}. 

Existing revocation mechanisms typically introduce centralised mechanics to handle revocations~\cite{alpar2017irma,tobin2016inevitable} or require Proof-of-Work blockchains~\citep{lundkvist_uport}. Both methods have limitations. First, relying on a centralised infrastructure may lead to censorship or privacy issues~\citep{sovrin_2017}. Second, blockchains suffer from privacy issues~\citep{yli2016current}, low throughput and limited flexibility~\cite{hughes2019blockchain}. Furthermore, they are prone to legislation limiting their use~\citep{xie2019china}.

The lack of a fully distributed revocation mechanism limits the mass deployment of Self-Sovereign Identities. It remains an open problem to design a revocation mechanism that does not depend on a central infrastructure or on third parties during verification and allows clients to independently verify credentials to prevent censorship.

We formulate the following problem description. First, an Issuer revokes their attestation for a credential. This revocation must be made apparent to all clients that may verify the credential and acknowledge the Issuer as an attestor. All clients are interconnected by a network overlay that they establish and maintain, and in which they have equal permissions. 
As direct communication with an Issuer or reliance on centralised infrastructure for verification goes against the principles of Self-Sovereign Identity, the propagation of revocations must be decentralised. Furthermore, neither an Issuer nor a receiving party can be expected to be online at all times, yet, all revocations are to be spread across the network in order to reach Verifiers. Furthermore, as all clients have equal permissions, they have to be able to individually decide whether or not to accept revocations.

\section{Related Work} 
\label{sec:related_work}
\autoref{tab:related_works} summarises our analysis of the revocation state-of-the-art in identity systems, and precises their respective limitations. This table also compares our revocation mechanism (see \autoref{sec:revocation_design}) to the related work. We consider the following characteristics: maturity of the solution, network operator requirement, availability of offline revocations, reliance on interactions with authorities (i.e., Issuers in the case of SSI), possibility of offline verification of revocations, presence of single points of failure (SPOFs), and accuracy of the verification mechanism (e.g. false positives or false negatives). As one can see, our solution qualitatively outperforms existing solutions. We note that blockchains allow for the realisation of distributed revocation. However, they suffer from obstacles such as privacy and security issues and low throughput~\cite{hughes2019blockchain}. 

As our key contribution addresses revocation, we focus on related work that discusses this topic. We found out that revocation in Self-Sovereign Identity is not widely discussed in academia, and as such, we selected works that address distributed revocation in a broader context. We organise the related works in groups that focus on the revocation of (SSI) credentials, certificates, and nodes.

Sovrin~\cite{tobin2016inevitable} and IRMA~\cite{alpar2017irma} propose the usage of cryptographic accumulators for \textit{revocation of (SSI) credentials} based on the works of Camenish et al.~\citep{camenisch2002dynamic}. A cryptographic accumulator is a probabilistic data structure that allows a large set of values to accumulate into a short witness value that can then be used to prove certain membership operations (e.g. inclusion checks). In the aforementioned solutions, a subject provides a proof of non-revocability of their credentials through this witness value. A verifier can then check this proof using the witness value, which is published on the blockchain. Sovrin does not allow for offline verification of credentials as both the subject and the verifier are required to retrieve the latest witness value during the verification of a credential. Similarly, IRMA does not allow offline verification because communication with its infrastructure is required. Furthermore, cryptographic accumulators can be computationally expensive to the extent that it is discouraged to use them for each verification in IRMA~\citep{alpar2017irma} and their probabilistic nature is prone to false positives.
Veramo (uPort)~\cite{lundkvist_uport} uses a single Ethereum smart contract for marking attestations as revoked. The usage of the Ethereum blockchain requires synchronisation of blocks in order to guarantee certainty on stored revocations. Furthermore, a single smart contract introduces a security risk~\citep{praitheeshan2019security}. Xu et al. use a blockchain to store legitimate subjects, indirectly disallowing access for revoked subjects in the SSI system~\cite{xu2020identity}. Updating this set is performed by the operators of the blockchain, which introduces centralised authorities. Abraham et al. propose the usage of a revocation list stored on a blockchain, on which consensus is reached through the nodes of the blockchain, maintained by operators~\cite{abraham_2020}. Offline verification is achieved through the storage of this revocation list. As the revocation list is not stored per authority, clients require full storage of this list, leading to storage overhead. We note that all revocations in an SSI system can grow up to gigabytes of storage, which hinders the deployment on devices with low memory (e.g. smartphones). Furthermore, the usage of a blockchain introduces further overhead as clients have to synchronise blocks. Stokkink et al. propose a fully distributed SSI system using direct peer-to-peer communication~\cite{stokkink2018deployment,stokkink2020truly}. They allow for three revocation mechanisms: i) linkage to a central revocation register, belonging to the Issuer; ii) the usage of distributed ledger technology; iii) the usage of short validity terms. The centralised approach opposes the sovereignty of the protocol, something that is acknowledged by the authors, blockchains again suffer from the aforementioned issues and short validity terms place too much power in the hands of Issuers. Speelman implements the approaches of~\cite{stokkink2018deployment,stokkink2020truly}, and additionally, proposes an active check as the main revocation method~\cite{speelman_2020}. Baars proposes the usage of smart contracts for storing revocations of attestations~\cite{Baars2016TowardsTechnology}, introducing security risks~\cite{praitheeshan2019security}.

Mechanisms for the \textit{revocation of PKI certificates} are present in traditional Public Key Infrastructures (PKIs) such as PKIX~\citep{pkix}. Broadly speaking, a PKI uses a Certificate Authority (CA) to publish a Certificate Revocation List (CRL), containing revoked certificates. In this structure, CAs are inherently central authorities, having relatively absolute power over revocations. These CAs, acting as trusted third parties, are central points of failure, suffer from MITM attacks, and are corruptable~\citep{allen2015decentralized}. 

PGP's web of trust~\cite{zimmermann1999} attempted to overcome this by handling revocation in a decentralised fashion, in which the revocation of keys was handled by the owner through revocation certificates. These certificates indicate that the key was compromised and should therefore no longer be used. However, PGP and its web of trust have been shown to be impractical~\citep{whitten1999johnny} and require central key servers.
Another alternative to PKI is the \textit{Decentralised Public Key Infrastructure} (DPKI)~\citep{allen2015decentralized,Fromknecht2014ARetention}. DPKI proposes the usage of alternative storage solutions for storing revocations of public keys. The proposed solutions use blockchains and, thus, require synchronisation of blocks for verification, introducing overhead and possibly low throughput as discussed previously.

Laberteaux et al. discuss the revocation of PKI certificates in vehicular ad hoc networks (VANETS) through the distribution of CRLs~\cite{laberteaux2008security}. Distribution is handled through Road Side Units (RSUs), serving as specialised nodes propagating the CRLs, and through epidemic spread between vehicles. The revocations are stored in Bloom filters. Haas et al. build upon this work by guaranteeing a certain degree of privacy by using group signatures~\cite{chaum1991group} when requesting certificates from the CA~\cite{Haas2011EfficientDistribution}. However, the revocations are handled by a single CA and the reliance on Bloom filters introduces the possibility of false positives. Liau et al. propose the distribution of CRLs through direct peer updates, reducing the communication overhead caused by periodic CRL synchronisation~\cite{liau2005}. Signatures over CRLs allow nodes to build trust in others. However, direct peer updates may prove to be suboptimal in the case of highly adaptive networks such as that of mobile devices. Propescu et al. discuss the revocation of certificates based on the clustering of clients and probabilistic auditing for honesty of revocation distributors~\cite{propescu2003}. This auditing is probabilistic in order to reduce performance requirements, however, this allows for malicious nodes to possibly exist for quite some time.

Eschenauer and Gligor discuss the \textit{revocation of nodes} in distributed sensor networks~\cite{eschenauer2002key}. Revocation is handled by a single node serving as an authority, delegating revocations to regular sensor clients. We note that the introduction of a single authority goes against the principles of SSI. Lasla et al. discuss the revocation of malicious vehicles in Cooperative Intelligent Transportation Systems (CITS)~\cite{Lasla2018EfficientITS}. They use a blockchain for storing revocations through a distributed vehicle admission and revocation scheme. Again, we note that blockchains suffer from the aforementioned hurdles such as privacy and security issues.

\section{System \& Threat Model}
\label{sec:system_model}
We focus on  the revocation of attestations and related SSI interactions in an \textit{identity network}. An identity network is a peer-to-peer (P2P) network that implements the identity service that is maintained and used by its peers. The network can be openly joined by any peer (also called a client) at any time. We build upon the terminology of the W3C's DID to define the potential roles of peers~\cite{sporny_longley_chadwick_2019}. A peer can assume one or several roles among those of Subject, Issuer and Verifier: a Subject is a client that holds credentials; an Issuer attests to a claim and is able to revoke its attestation; a Verifier verifies credentials. Any two clients are assumed to eventually be able to directly communicate with each other. We assume that clients are always connected to a subset of other clients, which are called their neighbours, with which they exchange identity-related information. Moreover, clients are deemed not to be necessarily always online. However, when they are online they are reachable and can communicate with other peers. As such, nodes periodically exchange membership information with their neighbours, replace them, and exchange revocation information with them. 

Adversaries or malicious actors may be present in the network. We assume that adversaries do not aid in maintaining the health of the network (via the spread of revocations discussed in~\autoref{sec:revocation_design}). These actors are not able to drop arbitrary messages but are able to send fabricated messages (e.g., replayed) to other clients. These messages are discarded by honest nodes as they can be evaluated as invalid.

\section{Architecture \& Theoretical Analysis} 
\label{sec:revocation_design}
Our revocation mechanism overcomes the hurdle of interactivity with authorities whilst enabling offline verification. Clients do not require to be online during verification of a credential, they merely require occasional synchronisation of revoked attestations by communicating with other nodes in the network. This is achieved through three concepts.

\subsection{Trusted Issuers}
In real life, a person has (relatively speaking) a choice of whether to acknowledge a certain authority. Following this fashion, this is also possible in our revocation architecture: each client manages, as we coin, a \textit{Trusted Issuer Storage} (TIS). The TIS is a register containing the public keys of the Issuers that are trusted by a client. Each of these Issuers is referred to as a \textit{Trusted Issuer} (TI). Hence, a distinction is made by each client, individually, on the trusted authorities in the network. A client exclusively accepts the revocations made by their TIs. The results of acceptance are the storage of the revocations by the client and further propagation of the revocations in the network. This significantly reduces storage requirements under the assumption that a client is not interested in revocations made by an Issuer that physically resides far away. TIs distinguish themselves from traditional TTPs as they are only able to influence the legitimacy of attestations that they created. Furthermore, they only aid in the verification process and, thus, do not provide a definite answer. 

Issuers publish their revocations as sets containing the hashes of revoked credentials. These sets are uniquely identified using a label and subsequent sets only contain new revocations. Our implementation uses the SHA3-256 algorithm for hashing and an incremental integer for version labelling. The version label is unique for each Issuer, but not across revocation sets made by other Issuers. The Issuer signs its set of revocations and the label for authenticity. Hence, an issuer revokes its attestation by publishing the hash of the credential that the attestation belongs to. This counteracts the possibility for clients to hide a revoked attestation.

\subsection{Attestation Revocation List}
All received revocations are stored by a client for later reference in, what we coin, its \textit{Attestation Revocation List} (ARL). The ARL is a register holding the revocations made by the TIs it trusts. It is, similarly to the TIS, stored and managed by each client individually. In the ARL, revocations are grouped by the TI that revoked the attestation and by the unique version label that is assigned by the TI. This makes it possible for a client to store duplicate revocations if multiple TIs revoked their attestation for the same credential. This is per design, as it allows Verifiers to build more trust in rejecting a revoked credential. Furthermore, storing revocations per TI counteracts malicious TIs from revoking attestations made by other Issuers.

As the number of revocations can grow to a large amount, we use Bloom filters~\cite{bloom_1970} to speed up the verification process for Verifiers, whilst overcoming their probabilistic nature. All Verifiers, in addition to storing the attestations, add them to a Bloom filter. Using their filter, Verifiers first test whether an attestation belongs to the ARL, after which, upon success, the definitive search is performed (to handle possible false positives). Raya et al.~\cite{raya2007} discuss the benefits of Bloom filters in Certificate Revocation Lists, which can provide similar speed improvements for the ARL, as both require validation of whether an item is part of a set of revoked items.

Furthermore, we note that the ARL can be replaced exclusively by a probabilistic data structure. A client may choose to accept the probabilistic nature of a Bloom filter over an exact membership check. Such clients are not able to aid in the propagation of the revocations, though the low memory requirements may sometimes prove to make the protocol suitable, e.g., for IoT devices. However, as a result, verification of credentials on the client may be affected by false positives. Whilst this does not explicitly impact security, it could lead to the false rejection of non-revoked credentials. As such, this is only suitable for Verifiers that expect to verify low numbers of credentials. For perspective: a Bloom filter of 907.24 KiB, using 10 hash functions, storing 100,000 items has a false positive probability of 1 in 1 billion.

\begin{figure*}[ht]
    \centering
    \begin{subfigure}[b]{0.325\textwidth}
        \centering
        \includegraphics[width=\textwidth]{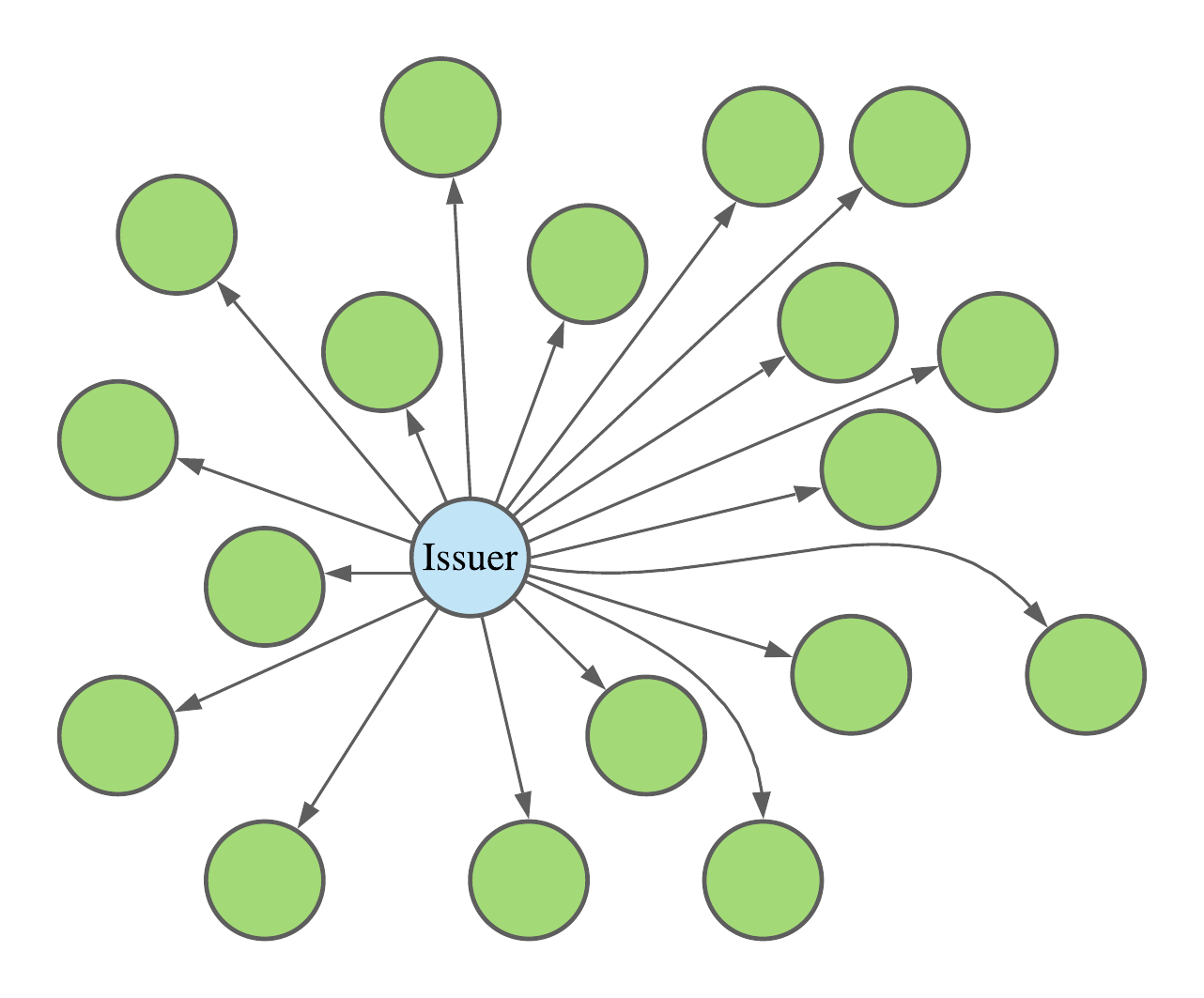}
        \caption[]%
        {{\small Centralised revocation propagation}}    
        \label{fig:hrm_0}
    \end{subfigure}
    \hfill
    \begin{subfigure}[b]{0.325\textwidth}  
        \centering 
        \includegraphics[width=\textwidth]{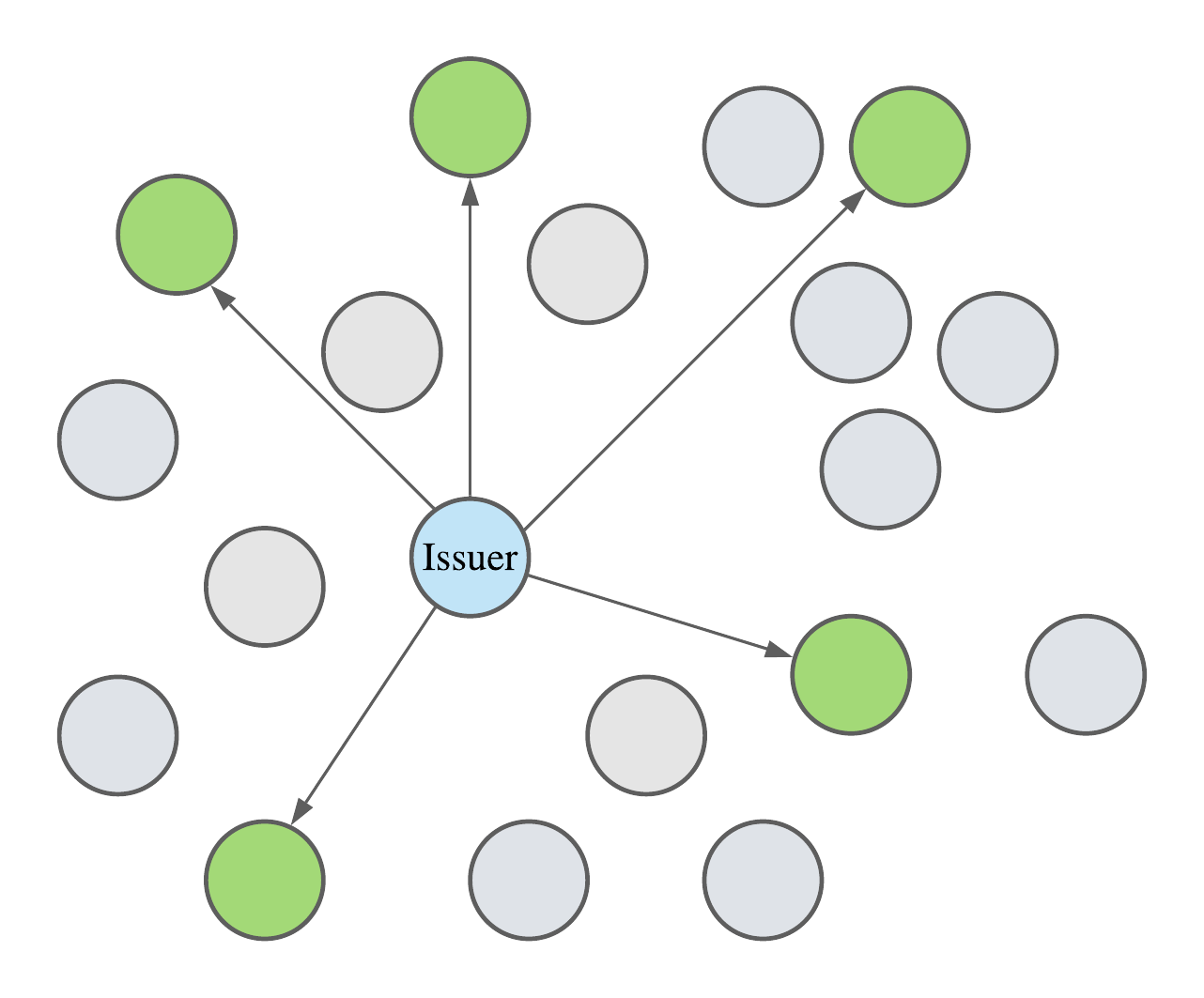}
        \caption[Indirect gossip propagation]%
        {{\small Direct propagation from Issuer}}    
        \label{fig:hrm_1}
    \end{subfigure}
    \hfill
    \begin{subfigure}[b]{0.325\textwidth}   
        \centering 
        \includegraphics[width=\textwidth]{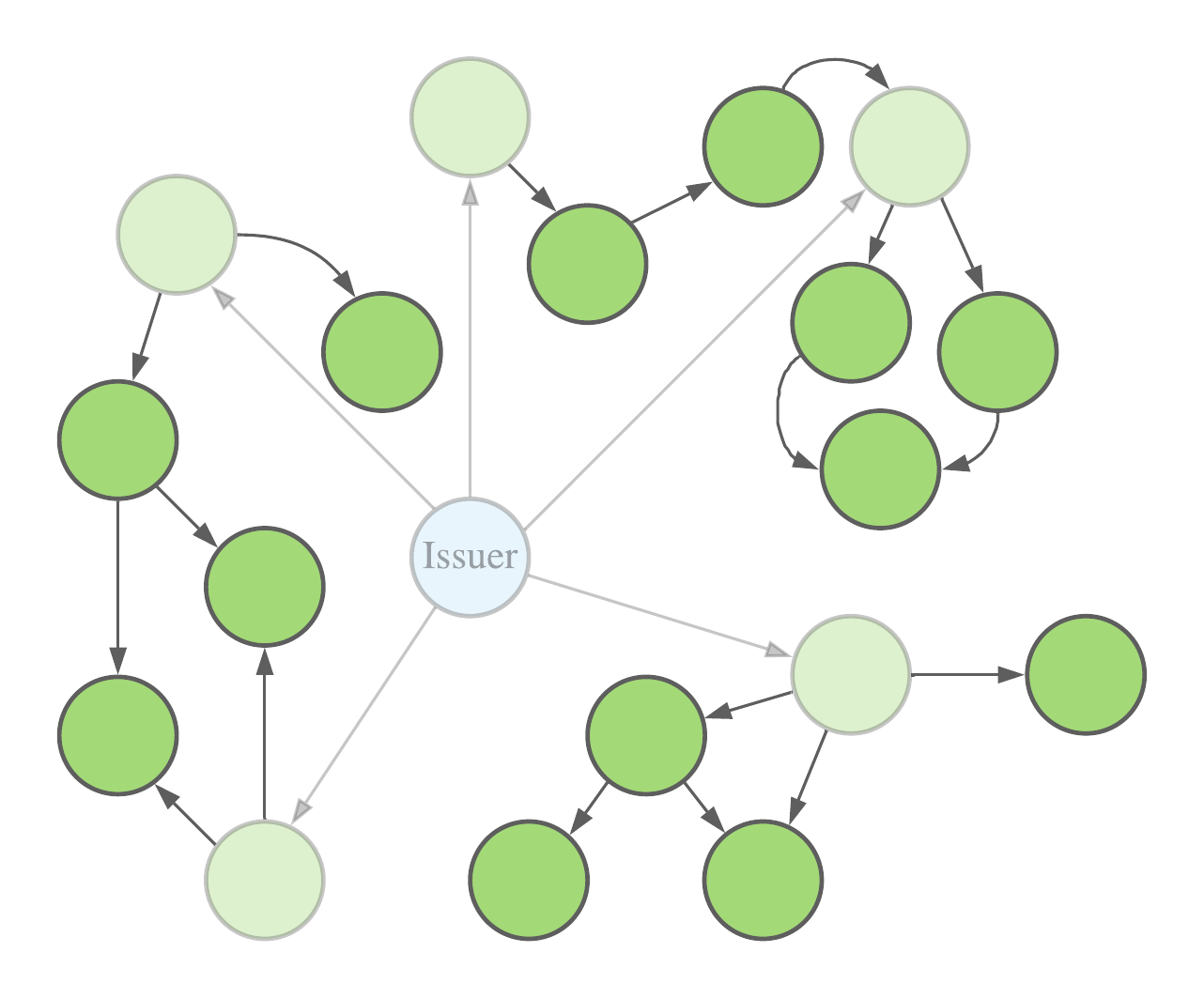}
        \caption[]%
        {\small Indirect propagation through gossip}    
        \label{fig:hrm_2}
    \end{subfigure}
    \caption{Fully distributed revocation} 
    \label{fig:hrm}
\end{figure*}

\subsection{Propagation}

In order to disseminate revocations, the architecture requires a protocol that ensures that information is spread across the entire network, whilst also ensuring that unavailable clients receive the information at a later instance. For this, we use a gossip protocol with static re-transmissions. Gossip protocols are communication protocols that allow for a periodic exchange of data with (random) peers~\citep{Kwiatkowska2008AnalysisPRISM}.

In order to maintain a low overhead and allow for selective revocation updates, revocations are propagated using advertisements. Gossiping nodes advertise their known revocations, after which a receiving node is able to selectively request revocations. Advertisements are structured as key-value pairs of the digest of a TI's public key and the latest version known by the gossiping client. A receiving client requests updates by sending back key-value pairs of the digests of each TI's public key it is interested in and the lowest version it is missing. The gossiping client then sends back all the requested revocations. Authenticity is guaranteed using the signatures. Based on the publication dates attached to revocations, clients are able to ignore revocations, optimising storage usage as they may no longer be relevant in the system due to expired validity terms of the corresponding attestations. 

\autoref{fig:hrm} illustrates the communication of revocations from an initial Issuer to a set of clients. \autoref{fig:hrm_0} portrays the traditional client-server model in which all revocations are received by the clients through a single central party, indicated by the green-coloured clients. This system is limited by the requirement of a direct link between all nodes and the central Issuer. A central Issuer may introduce security and availability issues and possibly leads to censorship.

Our gossip-based approach utilises the structure visible in \autoref{fig:hrm_1} and \autoref{fig:hrm_2}. \autoref{fig:hrm_1} portrays the initial gossip from an Issuer to a set of neighbours. In this instance, all clients are honest, acknowledge the Issuer, and are online during the propagation window (the time at which the Issuer has its gossip iteration). After which, the clients continue to propagate the revocations, in the same fashion, eventually spanning across the network. This is depicted in~\autoref{fig:hrm_2}. Thus, after the initial gossip by the Issuer, no further interaction with said Issuer is required for the propagation of the revocations. This figure also shows relays to clients that are already gossiped to, indicating that clients have multiple opportunities to receive the latest revocations. 

Our algorithm has a worst-case runtime of $\mathcal{O}(n)$, with $n$ being the network size, as in the worst case a single node updates the entire network. However, it is expected to be logarithmic with respect to the number of nodes, as each node that has received the latest revocations from a TI can gossip to the remaining uninformed nodes, speeding up the propagation time for the remaining nodes. Furthermore, not the entire network is interested in all revocations.

As becomes apparent from this description, dishonest nodes pose no large threat to the propagation of revocations. They could introduce a slight delay due to fewer clients gossiping information or due to the spread of fabricated revocations, which will be discovered by receiving clients. We do note that depending on the network topology, Eclipse attacks~\cite{castro2002secure} are a possibility that can be circumvented through direct connections with Issuers wherever possible.

\section{Algorithms \& Simulation} 
\label{sec:algs}
In order to realise the proposed revocation mechanism discussed in~\autoref{sec:revocation_design}, each client in the network runs three algorithms. A gossiping client runs~\autoref{alg:rev_gossip}, which enables the periodic advertisement of revocations. First, a random subset of peers is generated using the node selection function (line 2). This function generates a set of peers from the neighbours known by the client. Next, an advertisement is gossiped to each of these peers (lines 3-4). Finally, the gossiping client awaits the start of the next gossip interval (line 5). An advertisement consists of pairs of Issuer public keys and their latest known version of revocations.

A node receiving an advertisement runs~\autoref{alg:rev_req_update}. In this procedure, the node verifies whether any TI is present in the advertisement (lines 2-3). Then it verifies, using the \texttt{FindMissingVersion} routine, whether it misses or is uninformed about revocation versions belonging to the Trusted Issuer (lines 4-5). More specifically, \texttt{FindMissingVersion} determines on an advertisement containing the revocation version $v_i$ part of revocations made by TI $a_i$ with public key $pk_i$ whether $\exists (v_j,pk_i) \in \mathcal{ARL}$ such that $\forall (v_k,pk_i) \in \mathcal{ARL}$ it holds that $(v_j \geq v_k \wedge v_j < v_i)~\vee~(v_{j+1} \notin \mathcal{ARL} \wedge v_{j+1} < v_i)$. If this is the case, an update is requested from the gossiping client for the respective Issuer and the lowest missing version (line 6). The advertising node verifies whether it advertised to the node recently and sends the revocations.

Following the reception of requested revocations, a node executes~\autoref{alg:rev_update}. This procedure verifies the relevance of the revocations (line 1) and their validity (line 2). This validity check is performed by verifying the attached signature over the revocations and their version using the public key of the TI. Finally, the revocations are stored in the ARL (line 3).

\subsection{Simulation}
The analysis of the mechanism is two-fold. Firstly, we discuss a simulation showcasing scalability amongst a relatively high number of clients using the aforementioned algorithms. Secondly, we showcase analysis through the deployment on smartphones in~\autoref{sec:analysis}. The simulation was performed on a system with an Intel i7-6700HQ CPU clocked at 2.60 GHz and 16 GB of RAM.

\setlength{\textfloatsep}{8pt}
\setlength{\floatsep}{5pt}
\setlength{\intextsep}{5pt}

\IncMargin{1em}
\begin{algorithm}
\caption{Revocation advertisement gossip}\label{alg:rev_gossip}
    \SetKwData{Left}{left}\SetKwData{This}{this}\SetKwData{Up}{up}
    \SetKwFunction{Union}{Union}\SetKwFunction{FindCompress}{FindCompress}
    \SetKwData{In}{\textbf{in}}
    \SetKwData{Authority}{Authority}
    \SetKwData{Version}{Version}
    \SetKwData{Vloc}{$v_{local}$}
    \SetKwData{Client}{Client}
    \SetKwData{True}{True}
    
    \SetKwFunction{Fin}{FindMissingVersion}
    \SetKwFunction{Req}{RequestUpdate}
    \SetKwFunction{Gossip}{GossipRevocations}
    \SetKwFunction{Wait}{Wait}
    \SetKwFunction{Fin}{FindMissingVersion}
    \SetKwFunction{Sel}{SelectPeers}
    
    \SetKwInOut{Input}{input}\SetKwInOut{Output}{output}
    
        \Input{\Desc{$\mathcal{C}$}{Set of neighbours}\\ 
            \Desc{$\mathcal{A}$}{Set of known Issuer-version pairs} \\
            \Desc{$t_g$}{Gossip interval}\\
            \Desc{$n_g$}{Gossip amount}
        }
        \Output{
        Revocation advertisements
        }
        \BlankLine
        
        \While{\True}{
            $\mathcal{C}_g \leftarrow$ \Sel($\mathcal{C}, n_g$)\;
            \ForEach{$c_i \in \mathcal{C}_g$}{
                \Gossip{$c_i,\mathcal{A}$}\;
            }
            \Wait{$t_g$}\;
        }
        
\end{algorithm}\DecMargin{1em}

\IncMargin{1em}
\begin{algorithm}
    \SetKwData{Left}{left}\SetKwData{This}{this}\SetKwData{Up}{up}
    \SetKwFunction{Union}{Union}\SetKwFunction{FindCompress}{FindCompress}

    \SetKwData{In}{\textbf{in}}
    \SetKwData{Authority}{Issuer}
    \SetKwData{Version}{Version}
    \SetKwData{Vloc}{$v_{local}$}
    \SetKwData{Client}{Client}
    
    \SetKwFunction{Fin}{FindMissingVersion}
    \SetKwFunction{Req}{RequestUpdate}
    
    \SetKwInOut{Input}{input}\SetKwInOut{Output}{output}
    
    \Input{
        \Desc{$\mathcal{A}$}{Set of known Issuer-version pairs}
     }
    \Output{Revocation update request}
    \BlankLine
    \emph{On reception of $\mathcal{A}$ by \Client $c_i$}\;
    \For{\Authority $a_i$, \Version $v_j$ \In $\mathcal{A}$}{
        \If{$a_i \in \mathcal{TIS}$}{
            \Vloc $\leftarrow$ \Fin{$a_i$}\; 
            \If{\Vloc $< v_j$}{
                \Req{$c_i$,$a_i$,$v_{local}$}\;
            }
        }
    }

    \caption{Revocation update request procedure}\label{alg:rev_req_update}
\end{algorithm}\DecMargin{1em}

\IncMargin{1em}
\begin{algorithm}
    \SetKwData{Left}{left}\SetKwData{This}{this}\SetKwData{Up}{up}
    \SetKwFunction{Union}{Union}\SetKwFunction{FindCompress}{FindCompress}
    
    \SetKwData{In}{\textbf{in}}
    \SetKwData{Authority}{Issuer}
    \SetKwData{Version}{Version}
    \SetKwData{Vloc}{$v_{local}$}
    \SetKwData{Client}{Client}
    
    \SetKwData{True}{True}
    
    \SetKwFunction{Fin}{FindMissingVersion}
    \SetKwFunction{Req}{RequestUpdate}
    \SetKwFunction{Gossip}{GossipRevocations}
    
    \SetKwFunction{Wait}{Wait}
    \SetKwFunction{Fin}{FindMissingVersion}
    
    \SetKwFunction{Sel}{SelectPeers}
    
    \SetKwFunction{Ver}{Verify}
    
    \SetKwInOut{Input}{input}\SetKwInOut{Output}{output}
    
    \Input{\Desc{$R$}{Set of revocations}\\
    \Desc{$v_i$}{Revocations version}\\
    \Desc{$s_i$}{Signature}\\
    \Desc{$pk_i$}{Issuer public key}
    }
    \Output{$R \subseteq \mathcal{ARL}$}
    \BlankLine
    
    \eIf{\Authority $pk_i$ \In $\mathcal{TIS}$}{
        \eIf{\Ver{$pk_i$, $s_i$, $v_i|R$}}{
            $\mathcal{AR}L \leftarrow \mathcal{ARL} \cup R$\;
        }
        {
            $\bot$
        }
    }
    {
        $\bot$
    }
    \caption{Revocation reception}\label{alg:rev_update}
\end{algorithm}\DecMargin{1em}

\begin{figure}[t]
    \centering
    \includegraphics[width=0.9\linewidth]{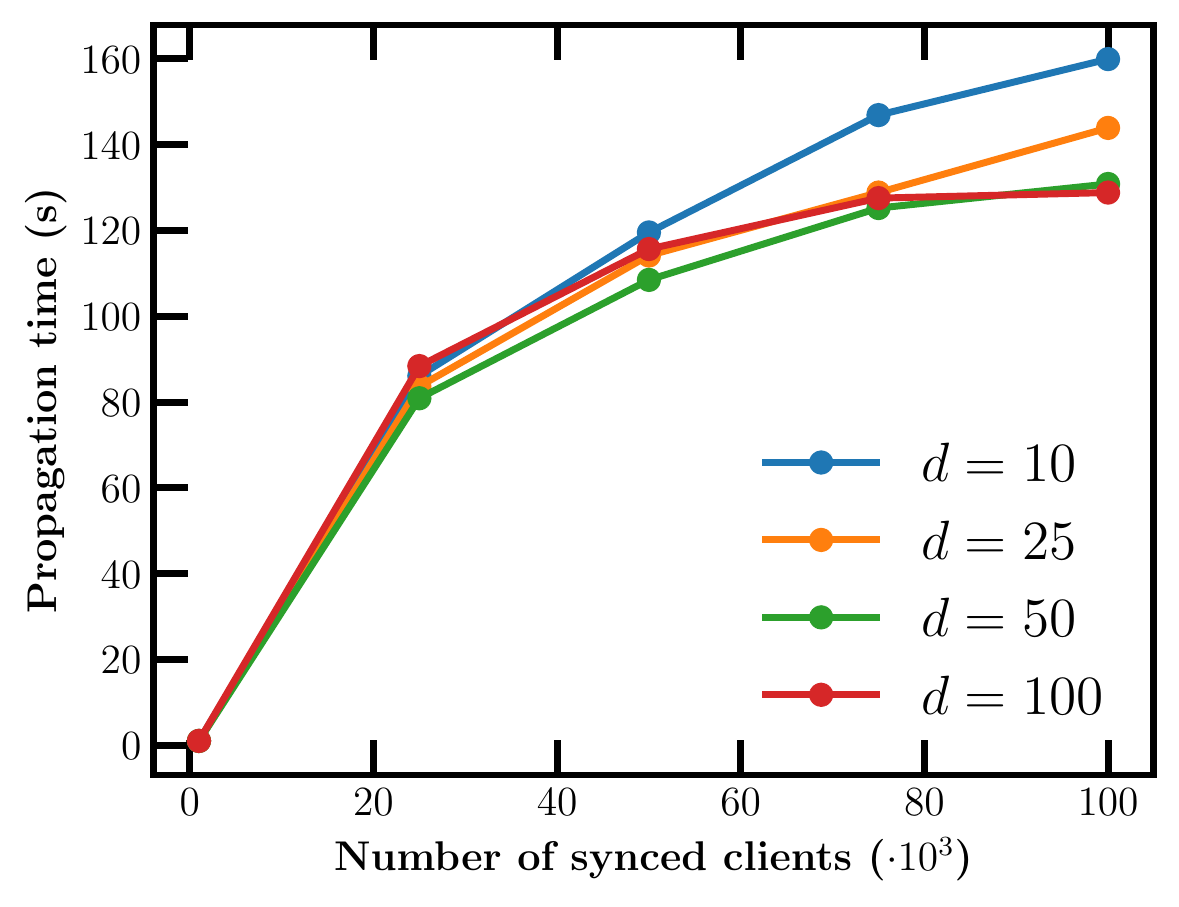}
    \caption{Revocation scaling}
    \label{fig:sim_new}
\end{figure}

The simulation has been performed through the mimicking of gossip of 340,000 revocations between clients released by a single client serving as a revoking Issuer. Each client runs the three algorithms and acknowledges the Issuer as a TI. The measurements of the simulation were gathered by simulating the execution time of the algorithm until full propagation of the revocations across a simulated network comprised of 100,000 nodes. The network is simulated using a weighted regular graph with differing degrees. A uniformly distributed weight between 0 and 20ms is introduced per link to simulate the impact of network latency, based on the global average reported by~\cite{speedtest_2021}. Furthermore, the cost associated with the transfer of 340,000 SHA3-256 hashes of 32 bytes is simulated using the global average upload speed of around 65 Mbps~\citep{speedtest_2021}. The revocations are released on $t=0$ by the Issuer. Each simulation is repeated 5 times. Due to multiple paths existing to each node, a client has multiple opportunities to receive revocations. Revocations are gossiped to all neighbours on reception.

\subsection{Simulation Results}
The averages of the timings are visible in~\autoref{fig:sim_new}. As expected, increasing the number of neighbours $d$ per client leads to lower propagation times. The simulation using $d=100$, however, shows a small decrease in performance when operating under fewer clients. Nonetheless, at 100,000 clients, this simulation leads to the lowest propagation time as expected. The observed decrease in performance, under fewer clients, can be explained by the overhead introduced by gossiping to 100 clients upon receiving new revocations. The results seem to indicate that the propagation increases logarithmically with respect to the number of clients and that the number of neighbours should be bound, not only due to performance limitations but also due to overhead in communication introduced by advertising to a larger number of clients. Overall, the simulation portrays that the algorithm is able to achieve realistic timings, taking up to 160 seconds to achieve propagation. In blockchain solutions, the propagation time depends on when the next block containing the revocations is mined, where the average block time on e.g. Bitcoin is 10 minutes plus transaction fees.

\section{Implementation \& Performance Analysis} 
\label{sec:analysis}
Sections \ref{sec:revocation_design} \& \ref{sec:algs} presented a novel fully distributed revocation algorithm with offline verification capabilities for Self-Sovereign Identity systems. Based on this architecture, two implementations have been made using a pioneering phone-to-phone infrastructure we developed previously\footnote{Official (Python) documentation: \url{https://py-ipv8.readthedocs.io/en/latest/}}~\citep{Halkes2011UDPWild,ZeilTech}. This serverless Web3 fabric allows for direct client-to-client communication, enabling a fully distributed infrastructure at the core of the solution. Using this implementation, we wrote an application for Android, backed by the Dutch National Office for Identity Data (RvIG)~\citep{RvIG2020DeIDee}, showcasing the feasibility on smartphones. The resulting implementations can be found in our public repositories\footnote{Infrastructure: \url{https://github.com/Tribler/kotlin-ipv8}}\footnote{Android application: \url{https://github.com/Tribler/trustchain-superapp}}.
 \smallskip
The analysis of the implementation has been performed in a test setup measuring the time required to gossip revocations between an Issuer and three Verifier clients running on smartphones. For revocations, we generated a dataset of 1 million revoked 32 bytes SHA3-256 hashes, a format used by the implementation. Revocations were split up into sets of 1000 in order to minimise the impact of a single packet loss. For the default parameters, the gossip-interval $t_{g}$ was set to $100$ ms in order to maximise the throughput of gossip. The number of selected peers $n_g$ was set to 5 as our phone-to-phone infrastructure uses 20 simultaneous connections per default. However, due to the network size of the test setup, the parameters are of minor impact.

\begin{table}[h]
\centering
\caption{}
\label{tab:phones}
\resizebox{0.7\linewidth}{!}{%
\begin{tabular}{|l|c|}
\hline
\textbf{Phone}     & \textbf{Propagation time (s)} \\ \hline
Galaxy s10         & 1066                                             \\ \hline
Pixel 2 XL (emul.) & 903                                              \\ \hline
Pixel 4 (emul.)    & 801                                           \\ \hline
\end{tabular}%
}
\end{table}

\subsection{Revocation Amount}
\autoref{tab:phones} showcases the revocation scaling in a system of 1 client gossiping revocations and 3 clients receiving revocations. Our results indicate that the propagation time scales linearly with respect to the number of revocations. One million revocations take up to 1066 seconds or just under 18 minutes. As this can be deemed more than 4 years' worth of revocations~\citep{proeftuin_2015} in the Netherlands, we deem this scalability usable as the propagation is expected to grow logarithmic with respect to the number of clients in larger networks.

Compared to the simulations discussed in~\autoref{sec:algs}, the performance is worse. We note that this can be explained mostly due to communication overhead caused by UDP packet splitting. The tremendous amount of packets led to many packet drops, in turn leading to the loss of specific revocation versions. As the reference implementation naively provides the gossiping client with a lower bound of missing versions, the additional network traffic of already gossiped versions causes additional packet losses. This snowballing effect worsens the performance of the algorithm. As such, the investigation of other network protocols or more sophisticated handling of packet loss can prove to significantly improve performance. However, the achieved performance can be deemed usable.

\section{Conclusion} 
\label{sec:conclusion}
This paper addresses revocation in Self-Sovereign Identity systems. We deem revocation to be the last remaining open issue for SSI to become a feasible contender for the next generation of identity systems. We proposed the first fully distributed revocation mechanism requiring no interactivity with any central parties, whilst adhering to the principles of the SSI paradigm. Revocations are propagated through the network using a gossip-based protocol, in which the acknowledgement of revocations is up to the discretion of Verifiers. Offline verification is enabled through local storage and no dependency on revoking Issuers. The feasibility of this revocation mechanism has been validated using a fully distributed SSI implementation. Our results show that fully distributed SSI is feasible on modern smartphones and that this is a promising direction to further explore. We conclude that our proposed architecture is a sizeable step towards placing identity back into the hands of the citizens.

\bibliographystyle{IEEEtranN}
\bibliography{main}

\end{document}